\documentclass[aps,prl,twocolumn,a4paper,byrevtex,superscriptaddress,showpacs,showkeys,bibnotes,longbibliography]{revtex4-1}
\usepackage{microtype}
\usepackage{graphicx}
\usepackage{amsmath}
\usepackage{amssymb}
\usepackage{color}
\usepackage[colorlinks,linkcolor=blue,urlcolor=blue,citecolor=blue]{hyperref}

\newcommand{\be}{\begin{equation}}
\newcommand{\ee}{\end{equation}}
\newcommand{\ra}{\rightarrow}
\newcommand{\idf}{1\!\! 1}

\newcommand{\cL}{\mathcal{L}}
\newcommand{\cH}{\mathcal{H}}

\begin{document}
\title{A minimal model of dynamical phase transition}

\author{Pelerine Tsobgni Nyawo}
\email{tsobgnipelerine@gmail.com}
\affiliation{Department of Physics, University of Stellenbosch, Stellenbosch 7600, South Africa}
\affiliation{\mbox{Institute of Theoretical Physics, Department of Physics, University of Stellenbosch, Stellenbosch 7600, South Africa}}

\author{Hugo Touchette}
\email{htouchette@sun.ac.za, htouchet@alum.mit.edu}
\affiliation{National Institute for Theoretical Physics (NITheP), Stellenbosch 7600, South Africa}
\affiliation{\mbox{Institute of Theoretical Physics, Department of Physics, University of Stellenbosch, Stellenbosch 7600, South Africa}}

\begin{abstract}
We calculate the large deviation functions characterizing the long-time fluctuations of the occupation of drifted Brownian motion and show that these functions have non-analytic points. This provides the first example of dynamical phase transition that appears in a simple, homogeneous Markov process without an additional low-noise, large-volume or hydrodynamic scaling limit.
\end{abstract}

\date{\today}


\maketitle

Dynamical phase transitions are phase transitions in the fluctuations of physical observables that give rise, similarly to equilibrium phase transitions, to non-analytic points in generalized potentials or large deviation functions characterizing the likelihood of fluctuations. Such transitions are known to arise in many physical systems and scaling limits, including the low-noise limit of diffusion equations modeling noise-perturbed dynamical systems \cite{freidlin1984,graham1984a,graham1985,graham1989,graham1995,bouchet2016}, thermodynamic-like limits of chaotic systems \cite{beck1993}, and the hydrodynamic limit of interacting particles systems, which corresponds, via the macroscopic fluctuation theory \cite{bertini2001,bertini2002,bertini2007}, to a low-noise limit \cite{bertini2010,bunin2012,bunin2013,aminov2014,baek2015}.
 
Similar transitions also appear in the long-time fluctuations of time-integrated quantities, such as Lyapunov exponents \cite{tailleur2007b,giardina2011,laffargue2013}, dynamical activities \cite{garrahan2007,garrahan2009,hedges2009,hooyberghs2010,jack2010,chandler2010,garrahan2010,garrahan2011,genway2012,ates2012,hickey2014,jack2014}, currents \cite{rakos2008,gorissen2009,gorissen2011,gorissen2012,hurtado2011b,espigares2013,hurtado2014,tsobgni2016,tizon2016}, and the entropy production \cite{gingrich2014,mehl2008,speck2012}, which now play a central role in studies of nonequilibrium processes. In this case, the large deviation functions are found to be smooth in the long-time limit; singularities start to appear only when a low-noise or a scaling (hydrodynamic, particle or mean-field) limit is taken in addition to the long-time limit \footnote{For quantum systems, this additional limit is often implicit; close inspection shows that it takes either the form of a mean-field limit or a low-noise limit \cite{garrahan2010,garrahan2011,genway2012,ates2012,hickey2014}.}, leading many to believe and claim that these additional limits are necessary for dynamical phase transitions to appear in Markov processes.

The study in \cite{harris2016} of non-homogeneous random walks that are reset in time has just shown, by a mapping to DNA models, that this is not always true -- a dynamical phase transition can occur in the long-time limit without a low-noise or scaling limit. Here, we present a simple, minimal model based on a one-dimensional and homogeneous diffusion process that confirms this. The process has no reset and so also shows that random resetting is not needed for such transitions to arise.

The process that we consider is the drifted Brownian motion, defined as
\be
X_t = \mu t+\sigma W_t,
\ee
where $\mu$ is the drift, $\sigma$ is the noise power, and $W_t$ is the simple, one-dimensional Brownian motion (BM) started at $W_0=0$. Physically, $X_t$ may represent the position of a small particle evolving in a fluid moving at slow, constant velocity or in a static fluid but with additional forces (created, e.g., by laser tweezers \cite{ashkin1997} or an AC trap \cite{cohen2005b}) that pull the particle at constant velocity. By analogy with electrical circuits perturbed by Nyquist noise \cite{zon2004a}, $X_t$ can also represent the charge dissipated in a resistor upon the application of a ramped voltage. In both cases, we are interested in the fluctuations of the \emph{fraction of time} that $X_t$ spends in some interval $[a,b]$ during a time $T$, which can be expressed as 
\be
\rho_T = \frac{1}{T}\int_0^T \idf_{[a,b]}(X_t)\, dt,
\ee
where $\idf_A(x)$ is the indicator function equal to $1$ if $x\in A$ and $0$ otherwise. This simple time-integrated quantity is also called the \emph{empirical measure} of $X_t$ in $[a,b]$ and is obviously such that $\rho_T\in [0,1]$, with $\rho_T=0$ corresponding to trajectories or paths of $X_t$ that never enter in $[a,b]$ and $\rho_T=1$ to paths that always stay in that interval.

For drifted BM or pure BM ($\mu=0$), $X_t$ is very unlikely to stay in any finite interval for a long time, so we must have $\rho_T\ra 0$ with probability 1 in the long-time limit $T\ra\infty$. This is confirmed by noticing that the density of $X_t$ gets flatter as time increases, and implies that the probability distribution $P(\rho_T=\rho)$ of $\rho_T$ concentrates on $\rho=0$ as $T\ra\infty$. 

From the theory of large deviations \cite{ellis1985,dembo1998,touchette2009}, we can infer that $P(\rho_T=\rho)$ scales with $T$ as
\be
P(\rho_T=\rho) \approx e^{-TI(\rho)},
\label{eqldp1}
\ee
so the concentration of the probability is in fact exponential in time. The function $I(\rho)$ defined by the limit
\be
I(\rho)=\lim_{T\ra\infty} - \frac{1}{T}\ln P(\rho_T=\rho)
\label{eqrf1}
\ee
is called the \emph{rate function} and characterizes, following (\ref{eqldp1}), the exponentially-small likelihood of observing small and large fluctuations of $\rho_T$ away from $0$. In practice, this function is most often not determined from the distribution of $\rho_T$ itself, which is usually unknown or very difficult to obtain, but indirectly from the so-called \emph{scaled cumulant generating function} (SCGF):
\be
\lambda(k)=\lim_{T\ra\infty} \frac{1}{T}\ln \langle e^{Tk\rho_T}\rangle.
\label{eqscgf1}
\ee
where $\langle\cdot\rangle$ denotes the expectation and $k$ is a real parameter conjugated to the occupation fluctuations. Under some general conditions (see \cite{touchette2009}), it can indeed be shown that $I(\rho)$ and $\lambda(k)$ are related by Legendre-Fenchel transform, so we can obtain the rate function as
\be
I(\rho)=\sup_{k}\{k\rho-\lambda(k)\}.
\label{eqlf1}
\ee

This large deviation formalism is the basis of the many recent studies cited before on the fluctuations of the activity, current and entropy production in many-particle Markov dynamics, random walks, and diffusion equations. Our goal now is calculate the SCGF and rate function of $\rho_T$ for drifted BM and to discuss the appearance of non-analytical points in these functions associated with dynamical phase transitions. The calculations follow the theory explained in \cite{angeletti2015}, so we will be brief. To simplify the discussion, we also present the calculations first for $\mu=0$ and then for $\mu\neq 0$. 

\begin{figure}[t]
\hspace*{-1in}
\includegraphics{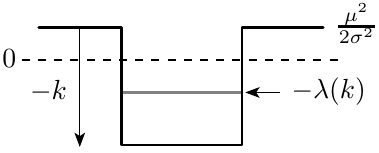}

\vspace*{-0.8in}
\includegraphics{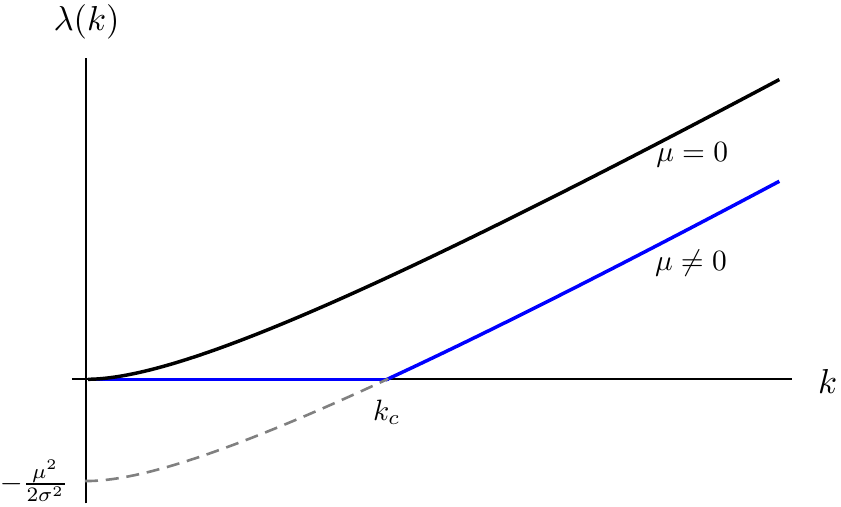}
\caption{(Color online) General form of the SCGF $\lambda(k)$ with (blue curve) and without drift (black curve). Inset: Equivalent quantum well problem determining $\lambda(k)$.}
\label{figscgf}
\end{figure}

\begin{figure}
\includegraphics{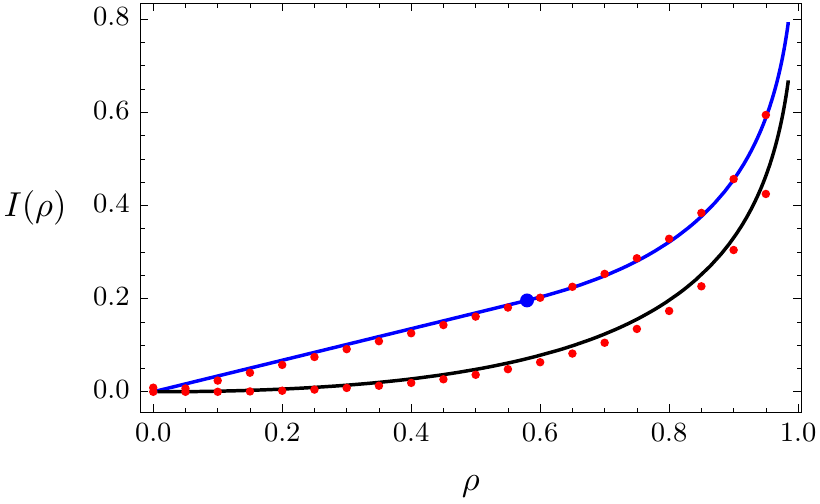}
\caption{(Color online) Rate function. Black curve: $I(\rho)$ obtained for $\mu=0$, $\sigma=1$ and $[a,b]=[-1,1]$. Blue curve: $I(\rho)$ obtained for $\mu=1/2$, $\sigma=1$, and the same occupation interval. The blue disk marks the phase transition point $\lambda'(k_c^+)=\rho_c$ below which $I(\rho)$ is linear with slope $k_c$. The red data points on each curve are the results of Monte Carlo simulations based on Euler integration of pure and drifted BM with $T=30$, $dt=0.1$, and $10^9$ samples.}
\label{figrf}
\end{figure}

\emph{Pure Brownian motion}: The SCGF of additive functionals such as $\rho_T$ is known to be given by the dominant eigenvalue of the so-called \emph{tilted generator} \cite{touchette2009}, which corresponds here to
\be
\cL_k  = \frac{\sigma^2}{2}\frac{d^2}{dx^2}+k\idf_{[a,b]}(x).
\ee
Up to a minus sign, this is just the quantum Hamiltonian associated with a finite well with depth $-k$ extending from $a$ to $b$, so that $\lambda(k)$ is simply minus the ground state energy of that well, as illustrated in the inset of Fig.~\ref{figscgf}. The behavior of the resulting $\lambda(k)$, which depends only on the difference $b-a$ and $\sigma$, is illustrated in the plot of Fig.~\ref{figscgf}. The main property to notice for our purpose is that $\lambda(k)$ is an analytic function in $k$ for all $k\geq 0$, since the ground state energy of the quantum well is itself known to be analytic with the depth of the potential well. This implies that the rate function $I(\rho)$, shown in Fig.~\ref{figrf}, is also analytic for all $\rho\in [0,1]$. Therefore, for pure BM there is no dynamical phase transition in the fluctuations of $\rho_T$ in the long-time limit.

\emph{Drifted Brownian motion}: For $\mu\neq 0$, the tilted generator becomes
\be
\cL_k  = \mu\frac{d}{dx}+\frac{\sigma^2}{2}\frac{d^2}{dx^2}+k\idf_{[a,b]}(x).
\ee
Although this operator is no longer self-adjoint, it is well known that it can be mapped to a Schr\"odinger-like operator $\cH_k$ by a unitary ``symmetrization'' transformation (see Case 2 in \cite{angeletti2015}) which gives here
\be
\cH_k=\frac{\sigma^2}{2}\frac{d^2}{dx^2}-\frac{\mu^2}{2\sigma^2} +k\idf_{[a,b]}(x).
\label{eqh1}
\ee
Consequently, $\lambda(k)$ is now determined by (minus) the ground state energy of the quantum well problem, shifted globally by a constant ``background'' energy $\mu^2/(2\sigma^2)$ (see inset of Fig.~\ref{figscgf}). This ``quantum'' eigenvalue, however, cannot give the complete SCGF because it becomes negative for $k<k_c$, as shown in Fig.~\ref{figscgf}. Here, $\lambda(k)$ must be positive, since it is convex, $\lambda(0)=0$, and
\be
\lambda'(0)=\lim_{T\ra\infty} \langle \rho_T\rangle=0.
\ee

To complete the analysis, we have to note that $\lambda=0$ is another possible eigenvalue of $\cL_k$. Its corresponding eigenfunction $r_k$ satisfying $\cL_k r_k=0$ does not decay to zero at infinity, but is such that $r_k(x)\ra 1$ as $k\ra 0$, which is the correct limit of the dominant eigenfunction \footnote{The eigenfunction associated with the quantum eigenvalue does not recover this limit, showing again that this solution does not give the SCGF as $k\ra 0$.}. Taking the maximum between this eigenvalue and the ``quantum'' eigenvalue above therefore yields
\be
\lambda(k)=\max \left\{0,\left.\lambda(k)\right|_{\mu=0}-\frac{\mu^2}{2\sigma^2}\right\},
\ee
The SCGF is thus a global shift of the $\mu=0$ solution only when the latter is positive.

With this result, we now see that $\lambda(k)$ is non-analytic at the critical point $k_c$ marking the transition from $\lambda=0$ to $\lambda\neq 0$. In fact, it is clear from Fig.~\ref{figscgf} that the first derivative of $\lambda(k)$ is discontinuous at this point, which means that the \emph{fluctuations of $\rho_T$ undergo a first-order phase transition when $\mu\neq 0$}: as $k$ crosses $k_c$, the occupation $\rho_k$ conjugated to $k$ via the Legendre relation $\rho_k=\lambda'(k)$ \cite{touchette2009} jumps from 0 (the long-time limit of $\rho_T$) to a strictly positive value $\rho_c=\lambda'(k_c^+)$, which converges to 1 as $|\mu|\ra\infty$. This applies regardless of the sign of $\mu$ since the global shift creating the non-analytic point is even in $\mu$.

The effect of this transition on the rate function is shown in Fig.~\ref{figrf}. The plateau and non-differentiable point of $\lambda(k)$ give rise, by the Legendre-Fenchel transform (\ref{eqlf1}), to a linear branch with slope $k_c$ which extends from $\rho=\lambda'(k_c^-)=0$ (so the minimum of $I(\rho)$ is actually a corner) to the critical occupation $\rho_c$. Beyond this point, $I(\rho)$ is simply the rate function obtained for $\mu=0$ shifted by the background $\mu^2/(2\sigma^2)$. In the limit $\mu\ra 0$, both $\rho_c$ and the shift go to $0$, thereby recovering the rate function of pure BM. This is confirmed in Fig.~\ref{figrf} by numerical results obtained by direct Monte Carlo sampling of $\rho_T$ \footnote{Simulations were done by sampling $10^9$ trajectories of BM and drifted BM, integrated using a Euler scheme with $T=30$ and $dt=0.1$. The rate function is obtained by applying (\ref{eqrf1}) without the limit and by shifting the minimal point, corresponding to the mean, at the origin.}, which also confirm our result for the SCGF.

The linear branch is reminiscent of ``coexisting phases'' observed in equilibrium first-order phase transitions (e.g., in the phase separation region of the liquid-vapor transition of water) and implies in our context that the fluctuations of $\rho_T$ are exponential in $\rho$ for $\rho\in[0,\rho_c]$. This region of fluctuations seems to be determined by paths of the drifted BM that reach a given occupation by ``controlling'' their final positions $x_T$, as observed in \cite{speck2012,chetrite2014,szavits2015}. Beyond $\rho_c$, the probability of $\rho_T$ is not exponential, since $I(\rho)$ is no longer linear. In this region of large occupations, fluctuations of $\rho_T$ are created by paths which stay in $[a,b]$ for a long time and which are therefore not affected by the drift. The probability of seeing such paths in drifted BM compared to pure BM can be computed using Girsanov's formula \cite{grigoriu2002}: it leads to an extra dominant factor $e^{-T\mu^2/(2\sigma^2)}$ as $T\ra\infty$, which explains the constant shift between the rate functions of drifted and pure BM.

We will provide a more detailed account of these two fluctuation regimes, together with the full solution of the dominant eigenfunction $r_k$, which shows a delocalization transition, in a future publication \footnote{P. Tsobgni Nyawo, H. Touchette, in preparation, 2016.} based on the recently-developed theory of Markov processes conditioned on large deviations \cite{chetrite2013,chetrite2014,chetrite2015}. At this point, the analytical calculation of $\lambda(k)$ is enough to show that the occupation fluctuations of drifted BM undergo a first-order dynamical phase transition, which is confirmed by numerical simulations. The transition does not involve a low-noise or large-system limit and also appears without broken ergodicity \cite{dinwoodie1993} or long-range (non-Markovian) correlations \cite{cavallaro2015}.

\newpage
This result and the simplicity of the model underlying it raise many interesting questions about the nature and conditions needed for the appearance of dynamical phase transitions. We suggest to conclude three particular questions that we see as important for future studies:

1. Can conditions known for equilibrium phase transitions (related, e.g., to dimensionality and the range of interactions) be reformulated for dynamical phase transitions? Some basic ideas related to that question can be found in \cite{rakos2008}. One way to map drifted BM or any other continuous-time processes to an ``equilibrium'' particle model is to view time as an extra space dimension and to relate the SCGF to a partition function.

2. What is the relation between the transition observed here and those observed in the weak-noise limit \cite{graham1984a,graham1985,graham1989,graham1995}, which have been re-discovered recently under the name ``Lagrangian phase transitions'' \cite{bertini2010}? This should be answered by comparing the variational representations of the SCGF and the rate function obtained with and without the low-noise limit \cite{chetrite2015}.

3. Does the phase transition of drifted BM disappear if we ``compactify'' this process on a ring or a closed interval? In other words, is the phase transition only a result of drifted BM evolving on an ``infinite'' state space (the real line)? It is commonly believed that the SCGF of compact, irreducible Markov processes must be analytic, but to our knowledge this has yet to be proved in general beyond the cases of finite Markov jump processes and Markov chains. Moreover, although $X_t$ is unbounded, the occupation observable $\rho_T$ for which the phase transition occurs is bounded. Finally, there are many models that evolve in infinite space and yet do not give rise to phase transitions (e.g., pure BM), so the issue of infinite versus finite spaces is a subtle one. 

\begin{acknowledgments}
We thank Cesare Nardini, Fr\'ed\'eric van Wijland, and Rapha\"el Chetrite for useful comments. P.T.N.\ is supported by a DAAD Scholarship. H.T.\ is supported by the National Research Foundation of South Africa (Grants no.\ 90322 and 96199) and Stellenbosch University (Project Funding for New Appointee).
\end{acknowledgments}

\bibliography{masterbib}

\end{document}